\documentclass[preprint,floats,aps,epsfig,nofootinbib,amssymb]{revtex4}
\usepackage{graphicx}
\usepackage{dcolumn}
\usepackage{bm}
\usepackage{mathrsfs}
\usepackage{slashed}
\usepackage{graphicx,color}
\usepackage{epsfig}
\usepackage{subfigure}
\usepackage{CJK}
\usepackage{xcolor}
\usepackage{amsmath}
\usepackage{amssymb}
\usepackage{caption2}
\usepackage{amsmath}
\usepackage[]{hyperref}

\begin{document}
\bibliographystyle{prsty}
\title{Search for
supersymmetric mesinos near production threshold in terms of the superflavor symmetry}
\author{Lian-Bao Jia$^1$}
\author{Xing-Dao Guo$^1$}
\author{Hong-Ying Jin$^2$}
\author{Xue-Qian Li$^1$}
\author{Ming-Gang Zhao$^1$}
\affiliation{1. School of Physics, Nankai University, Tianjin
300071, P. R. China \\
2. School of Physics, Zhejiang University, Hangzhou, Zhejiang Province, P. R. China\\}

\begin{abstract}

The supersymmetry (SUSY) may be one of the most favorable extensions of the standard model (SM), however, so far at LHC
no evidence of the SUSY particles were observed. An obvious question is whether they have already emerged, but escaped from our detection, or do not exist at all. We propose that the future ILC may provide sufficient energy and luminosity to produce SUSY particles as long as they are not too heavy. The superflavor symmetry associates production rates of the SUSY mesinos with that of regular mesons because both of them contain a heavy constituent and a light one. In this work, we estimate the production rate of SUSY mesinos near their production threshold and compare with  $B\bar B$ production. Our analysis indicates that if the SUSY mesinos with masses below $\sqrt s/2$ ($\sqrt s$ is the ILC energy) exist, they could be observed at future ILC or even the proposed CEPC in China.

\end{abstract}

\maketitle

\section{Introduction}

As is well known, the most important goal of high energy research is to look for new physics beyond
standard model (BSM), and SUSY may be the most favorable extension of the standard model(SM) because it can
reasonably explain the naturalness problem of Higgs and provide a dark matter candidate. Moreover,
its existence makes the strong, eletromagnetic and weak interactions to merge into one point
at the grand unification scale \cite{Langacker:1991an}. However, so far, at Tevatron and LHC, no SUSY particles
have ever been observed. One may wonder if the SUSY model is wrong or should be radically modified. Of course,
there is one more possibility that the SUSY particles have indeed been produced, but are not identified, namely
buried in the messy background at hadron colliders. Several authors \cite{Li:2013uma} notice this possibility and have tried to
reanalyze the LHC data and indicated the probability of misidentifying the SUSY particles.

In the minimal supersymmetric standard model (MSSM) and the modified SUSY models,
the scalar top quark has two mass eigenstates,
$\tilde{t}_1$ and $\tilde{t}_2$, and the lighter one ($\tilde{t}_1$) is assumed to be
the lightest squark.
Generally, it is believed that the lightest supersymmetric particle (LSP) is the
colorless neutralino $\tilde{\chi}_1^0$. The present results of the CMS and ATLAS Collaborations for searching scalar top quark
can be found in Ref. \cite{cms-stop,atlas-stop}, and it is noted that there is still possibility
for stop with mass of a few hundreds of GeV, e.g. there are windows:
$m(\tilde{t}_1)>200$ GeV with $m(\tilde{t}_1)-m(\tilde{\chi}_1^0)<m(W)$, and a heavier stop as
$m(W)<m(\tilde{t}_1)-m(\tilde{\chi}_1^0)<m(t)$. The literature suggests that
considering the 125 GeV Higgs boson observed at LHC, a sub-TeV stop could be allowed by the data \cite{Buckley:2012em,Barger:2012hr}.

It is also widely recognized that the hadron collider is a machine for discovery, whereas the electron-positron
collider is for precise measurement and unambiguous confirmation of discoveries.
As long as the SUSY theory or its modified versions are valid and
the stop mass is within the energy ranges of LHC and ILC, the stop
pair should be produced at those machines with sufficient detection efficiency. At the hadron colliders, the signals of the produced
SUSY particles might be buried in the messy background, so one may turn to the electron-positron collider to search for evidence of their existence.

In literature, it is suggested that the squark
$\tilde{t}_1$ is the next-to-lightest supersymmetric particle (NLSP).
If the mass of $\tilde{t}_1$ is not far away from that of LSP, its lifetime could be longer than $1/\Lambda_{QCD}$ \cite{Hikasa:1987db,Boehm:1999tr,Djouadi:2000bx,Djouadi:2001dx,Das:2004iz,Sarid:1999zx}, and it can attract a SM quark(anti-quark) to form a color singlet SUSY hadron~\cite{Farrar:1978xj,Dimopoulos:1996vz,Beenakker:1996ch,Sarid:1999zx,Kraan:2004tz}: the mesinos
after production. For SUSY mesinos consisting of $\tilde{t}_1$ and a heavy anti-quark $\bar{Q}$(Q=c,b), the fragmentation functions were calculable through
perturbative QCD, and they were studied by Chang et al. \cite{Chang:2006ns}. In their scheme,
to reliably determine the initial condition for the evolution differential equation,
the SM quark must be heavy so that perturbative QCD can apply. Obviously, the production rate for such processes is much suppressed. Whereas, if the SM constituent quark is light (u,d,s) the production rate might be greatly enhanced, however unfortunately then the non-perturbative QCD effects would be dominant, so that the perturbative computation
becomes not reliable. An alternative method for evaluating the production rate of such mesinos is needed.

In this work, we focus on the production rate of SUSY mesino which consists of
a heavy scalar quark and a light SM antiquark at $e^+e^-$ colliders. The production rate of a pair of SUSY squark-anti-squark at electron-positron
collider have been well calculated at the tree-level and loop-level (see, e.g. \cite{Arhrib:1994rr,Bartl:2000kw,Jimbo:2012qw}), thus the key point is how to calculate the hadronic matrix elements which are fully governed by the
non-perturbative QCD. Obviously, to directly evaluate the relevant hadronic matrix elements one needs to invoke concrete models.
In an interesting scheme we could associate the B-meson production
near its threshold which is well measured by CLEO~\cite{Artuso:2005xw}, Belle~\cite{Drutskoy:2006fg}, and BaBar~\cite{Aubert:2008ab} collaborations, with the production
of SUSY mesinos which may be obtained at ILC near their threshold by means of the superflavor symmetry \cite{Georgi:1990ak}.

The superflavor symmetry establishes a definite relation between the processes involving heavy mesinos and heavy mesons, where both the mesino and meson contain a heavy constituent
and a light quark(anti-quark). For the meson case the heavy constituent is a heavy quark(anti-quark)
of color-triplet(anti-triplet) fermion  $b(\bar b)$ or $c(\bar c)$, whereas for the SUSY mesino case
the heavy constituent is a color-triplet(anti-triplet) scalar. In our earlier work \cite{Guo:1994fn},
we supposed the heavy constituent to be a heavy diquark ($bb,\;bc$ or $cc$) whose inner structure
may manifest as a complicated form factor. Of course, it is more natural to apply the theory to the SUSY case
where the heavy constituent in the mesino is a point-like color-triplet (anti-triplet) squark (anti-squark).
Once we have the relation,
we can associate the production rate of the SUSY mesinos at ILC with the measured production rates of the B meson at the B-factories or LHCb.

In the heavy flavor mass limit, the QCD contribution in heavy
flavor hadron is independent of the heavy flavor's mass and spin. When we adopt the superflavor
symmetry to estimate the production rate of SUSY mesino, the measured B-meson production rate can be
obtained simultaneously.

In the ILC technical design report (volume II) \cite{Baer:2013cma}, the
top squark $\tilde{t}_1$ is expected to be found as long as $m_{\tilde{t}_1}\leq \sqrt{s}/2$.
At early stage, ILC will be running at
$\sqrt{s}=500$ GeV with luminosity $500$ fb$^{-1}$. In this stage, the $\tilde{t}_1$
mass will be determined to $1$ GeV and even $0.5$ GeV  accuracy \cite{Baer:2013cma,Keranen:2000py}.
Then its center of mass energy will be upgraded to $1$ TeV with luminosity $1000$ fb$^{-1}$.
At that energy scale, a SUSY particle with mass less than $0.5$ TeV could be found, and if considering
possible R-violation, even heavier SUSY particles might be observed.

The work is organized as follows: after this introduction where we explicitly introduce
our scheme, we formulate the cross sections for SUSY mesino $\tilde X$ and heavy SM meson $B$ in terms of superflavor symmetry in Sec. II.
In Sec. III, we present our numerical results along with all input parameters, and we
especially show how to associate the mesino production with B-meson at B-factories. The last section
is devoted to our conclusion and some discussions.

\section{The superflavor symmetry and the SUSY mesino production}

\subsection{The superflavor symmetry and its application}

Let us first have a brief review of the superflavor symmetry, and then focus on its application. Georgi and Wise extended the heavy quark's spin and flavor symmetry and introduced the superflavor symmetry \cite{Georgi:1990ak}. The superflavor symmetry relates the processes involving a heavy meson made of a heavy quark $h_v^+$ and a light anti-quark to a heavy fermion (mesino) made of a color triplet scalar $\chi_v$ (here we suppose it to be a squark) and a light color anti-triplet quark. The lagrangian of the heavy triplets with velocity $v$ is \cite{Georgi:1990ak}
\begin{eqnarray}
\mathscr{L}_v=\frac{1}{2}i(\bar{h}^+_v v_\mu\overleftrightarrow{D}^\mu{h^+_v}+2m_\chi\chi_v^\dag v_\mu\overleftrightarrow{D}^\mu\chi_v)\,.
\end{eqnarray}
Putting $h^+_v$ and $\chi_v$
altogether into a 5-column vector with a given velocity $v$, one has
\begin{eqnarray}
\Psi_v=\left(\begin{array}{c}
    h^+_v     \\
      \chi_v
      \end{array}\right).
\end{eqnarray}
Here one can write the wavefunctions of the meson and mesino consisting of
$h_v$ and $\chi_v$ as
\begin{eqnarray}
\Psi_H(v)=\left(\begin{array}{c}
    \sqrt{m_h}\gamma_5\frac{1}{2}(1-\rlap /v)     \\
      0
      \end{array}\right)
\end{eqnarray}
and
\begin{eqnarray}
\Psi_X(v)=\left(\begin{array}{c}
   0     \\
     \frac{u^T C}{\sqrt{2m_\chi}}
      \end{array}\right),
\end{eqnarray}
where $C$ is the charge conjugation operator and $u$ is the spinor wave
function of the $\chi$ bound state.

In the heavy quark effective theory (HQET)~\cite{Isgur:1989vq,Isgur:1989ed,Luke:1990eg},
for the transition of $b\to c$, gluons (or photon) are exchanged at t-channel and
the hadronic transition matrix element can be described by a unique Isgur-Wise function
$\xi(\omega)$ where $\omega={v}\cdot{ v'}$ is the recoil variable and ${v}$,
${v}'$ are the four-velocities of initial and final heavy mesons.  For the production
process, the gluon, photon or $Z_0$ (see in the following) is exchanged at s-channel and the kinematic region is different as ${v}\to -{v}$ \cite{Guo:1994fn}.
We need to generalize the Isgur-Wise function to the kinematic region of production, and some discussion about this situation was given in Ref. \cite{Guo:1994fn}.

From the matrix elements of meson and mesino given by Georgi and
Wise \cite{Georgi:1990ak}, the corresponding forms at pair production are
\begin{eqnarray}
\langle H(v')\bar H(v)|J^{\prime\mu}|0>=\langle H(v')\bar H(v)|\bar h \gamma^\mu h|0\rangle= \xi(-v\cdot v')m_h (v'-v)^\mu,\\
\langle X(v')\bar X(v)|J^{\mu}|0>=\langle X(v')\bar X(v)|i\chi^\dag\overleftrightarrow{\partial}^\mu\chi|0\rangle=\xi(-v\cdot v')\frac{1}{2} (v'-v)^\mu\bar u'v,
\end{eqnarray}
where $\xi(-v\cdot v')$ is the Isgur-Wise function, $\xi(1)=1$ at zero recoil point.

It is natural to apply the superflavor symmetry to SUSY hadron production.
In the heavy flavor mass limit, in high energy collisions, $b\bar b$ or stop pairs are produced, and then $b$ and $\bar b$ or $\tilde{t}_1$ and $\bar{\tilde t}_1$ hadronize into bound states by attracting antiquark(quark) from vacuum. The two different processes ($b\to$hadron and $\tilde t_1\to$ SUSY hadron) can be connected
by the superflavor symmetry. Obviously, a heavy quark fragments into a double heavy flavor meson (for example $b\to \bar B_c(b\bar c)$) is more
suppressed compared with a single heavy meson (for example $b\to \bar B_d(b\bar d)$) by a factor of $10^{-4} \sim 10^{-3}$ \cite{Chang:1991bp}.
The case of the SUSY hadron production is similar, i.e. production of mesino $\tilde t_1 \bar b(\bar c)$ is more suppressed than $\tilde t_1 \bar q\; (q=u,d,s)$.

A theoretical estimate shows that the so called  stoponium can be formed, and the binding energy is about 1-3 GeV \cite{Kim:2014yaa} which is much smaller
than the mass of stop and does not affect the phase space of the production.

Next we calculate the production rate of
$e^+e^-\rightarrow \tilde X\bar{\tilde X}$ near its threshold at ILC, whose low background makes it more advantageous over hadron colliders.

\subsection{Estimating the SUSY mesino production rate}

To predict the production rate of the SUSY mesinos near threshold, one could associate it with
the B-meson production near threshold. Indeed, we wish to use the data of the B-factory to predict
the production rates as
\begin{equation}
{\sigma^{theor}(e^+e^-\to \tilde X\bar{\tilde X})\over \sigma(e^+e^-\to \tilde X\bar{\tilde X})}\sim
{\sigma^{theor}(e^+e^-\to B\bar B)\over \sigma^{exp}(e^+e^-\to  B\bar B)},
\end{equation}
where the superscript "theor" means the theoretically predicted value and $\sigma^{exp}(e^+e^-\to  B\bar B)$ is the
measured value at B-factories and $\sigma(e^+e^-\to \tilde X\bar{\tilde X})$ is what we expect. The ratio of
$${\sigma^{theor}(e^+e^-\to \tilde X\bar{\tilde X})\over \sigma^{theor}(e^+e^-\to B\bar B)}$$ can be obtained in terms of
the superflavor symmetry, so that one can eventually obtain $\sigma(e^+e^-\to \tilde X\bar{\tilde X})$. In fact,
by the superflavor symmetry we can relate the matrix element $<\tilde X\bar{\tilde X}|J^{\mu}|0>$ to the matrix element
$<B\bar B|J^{\prime\mu}|0>$,  where $J^{\mu}$ and $J^{\prime\mu}$ are vector currents corresponding to squark-anti-sqaurk and
quark-anti-quark productions respectively.

However, there is a serious problem  that all the available data about the B-meson productions are not exactly
what we need, because the available data are from $e^+e^-\rightarrow \Upsilon(4s) /\Upsilon(5s) /\Upsilon(6s)\to B\bar B$,
namely via the $\Upsilon$ resonances. Instead, we need the data on the direct production of
$e^+e^-\to B\bar B$, i.e the contribution of the continuum spectrum near the threshold.
The total spectrum on $R_b$ (defined as $R_b={\sigma(e^+e^-\to hadrons)\over\sigma(e^+e^-\to\mu^+\mu^-)}$)
provided by experimentalists \cite{Aubert:2008ab} which is close to 0.3 also cannot be
used either.\footnote{For this point, we thank Dr. C.Z. Yuan of IHEP who told us that there are no such data about the continuum spectra available,
and also there is no an appropriate way to extract the continuum contribution from the data so far.}

Therefore, we can only associate the production rates of mesino with the B-meson production rates, but so far we cannot use the data
to make a definite prediction yet. However,
we expect our smart experimental colleagues to figure out an elegant way to extract the continuum contribution from the data or
directly measure it in the future (we believe that they will be able to do it), then we will obtain more accurate numerical results
of the mesino production rate near threshold.

Below we will derive the transition amplitudes and cross sections for the processes
$e^+e^-$ to $B\bar B$ and $\tilde X\bar{\tilde X}$, where $B$ and $\tilde X$ denote the meson and mesino
respectively. For the process $e^+e^-\rightarrow B\bar B$ at B factories, the collision energy $\sqrt s$
is much less than the mass of $Z_0$, thus the $Z_0$ contribution can be safely ignored.
By contrast, since in the process $e^+e^-\rightarrow \tilde X\bar{\tilde X}$,
$\sqrt s$ is larger than the mass of $Z_0$, the $Z_0$ contribution must also be included. The differential cross section for B-meson is
\begin{eqnarray}
&d\sigma(B \bar{B})= {1\over 8s_1}\displaystyle {\sum_{s_i, s_f}}
\big {|}\frac{-i}{3}e\langle B\bar B|\bar b\gamma^\mu b|0\rangle\frac{1}{s_1}
\langle 0|\bar e(-ie)\gamma_\mu e|e^+e^-\rangle \big {|}^2d\tilde v ,
\end{eqnarray}
where only photon contribution is taken into account, and for mesino it is
\begin{eqnarray}\begin{array}{rl}
d\sigma(\tilde X\bar{\tilde X})=&{1\over 8s_2}\displaystyle {\sum_{s_i,s_f}}
\big {|} \frac{2i}{3}e\langle\tilde X\bar{\tilde X}|\tilde t_1^\dag\overleftrightarrow{\partial}^\mu\tilde t_1|0\rangle
\frac{1}{s_2}\langle 0|\bar e(-ie)\gamma_\mu e|e^+e^-\rangle\\
&+g_{ t z}\langle\tilde X\bar{\tilde X}|\tilde t_1^\dag\overleftrightarrow{\partial}^\mu\tilde t_1|0\rangle
\frac{1}{s_2-m^2_Z}\langle 0|\bar e \gamma_\mu g_{ez} e|e^+e^-\rangle
\big {|}^2d\tilde v,
\end{array}\end{eqnarray}
where $g_{ t z}=\frac{ie}{\sin\theta_w\cos\theta_w}(\frac{1}{2}\cos^2\theta_{ t}-\frac{2}{3}\sin^2\theta_w)$
is the coupling constant between stop and $Z_0$ boson, $\theta_{t}$ in $g_{t z}$ is
the stop mixing angle \cite{Hikasa:1987db},
$\theta_w$ is Weinberg angle,
$g_{ez}=\frac{-ie}{\sin\theta_w\cos\theta_w}(\frac{1-\gamma_5}{4}-\sin^2\theta_w)$
is the coupling constant between electron and $Z_0$ boson,
$\sqrt{s_1}$ is the center of mass energy of B factory and $\sqrt{s_2}$ is
the center of mass energy of ILC. It is noted that $s_i$ is the spin projections of the
electron and position in the initial state and $s_f$ is the spin projections of the produced
B mesons or SUSY mesinos in the
final state and $d\tilde v$ is the corresponding final state phase space.

\begin{figure}
\centering
\begin{minipage}[!htbp]{0.41\textwidth}
\centering
\includegraphics[width=0.98\textwidth]{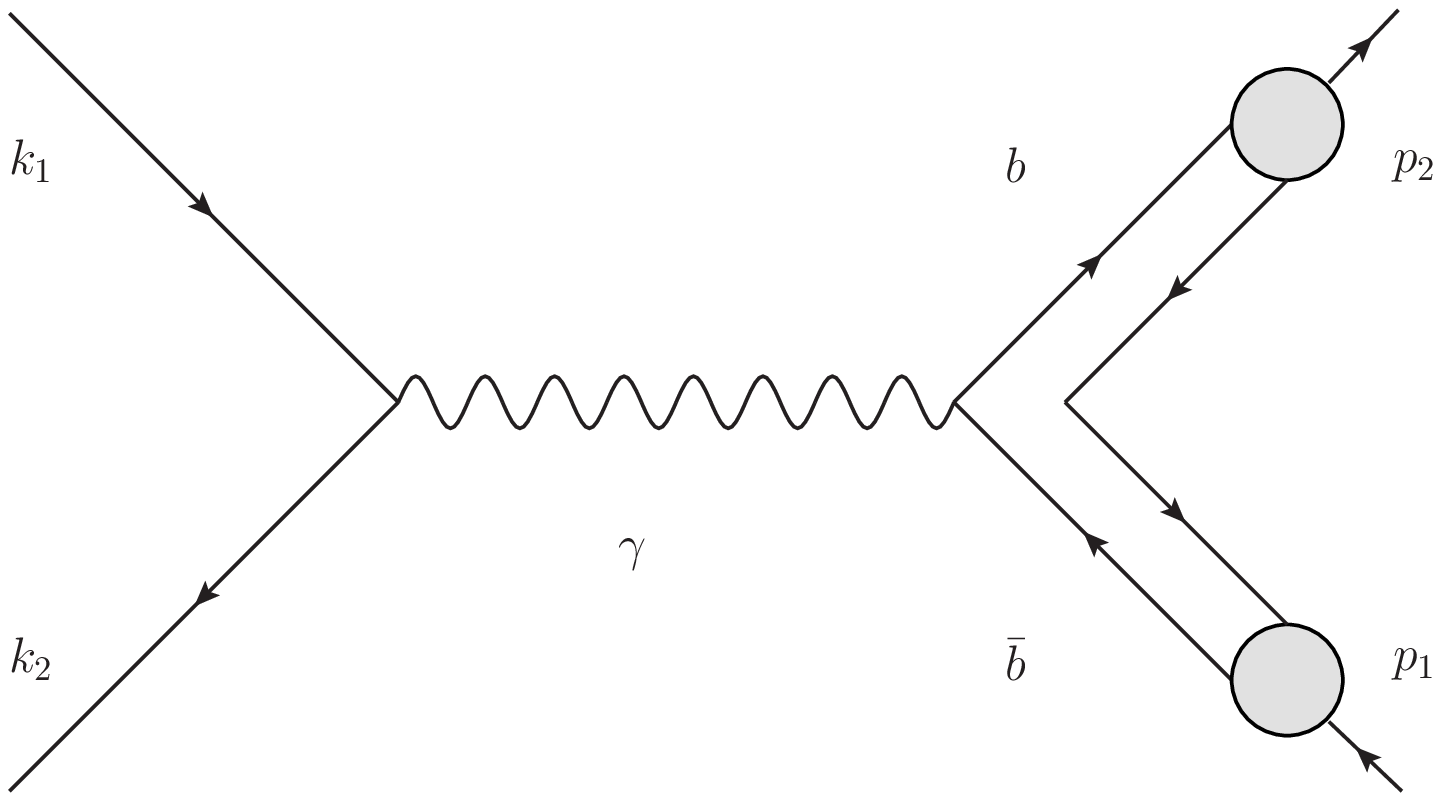}
\caption{The process of $e^+e^-\rightarrow B\bar B$.}
\label{bbmeson}
\end{minipage}
\begin{minipage}[!htbp]{0.4\textwidth}
\centering
\includegraphics[width=0.98\textwidth]{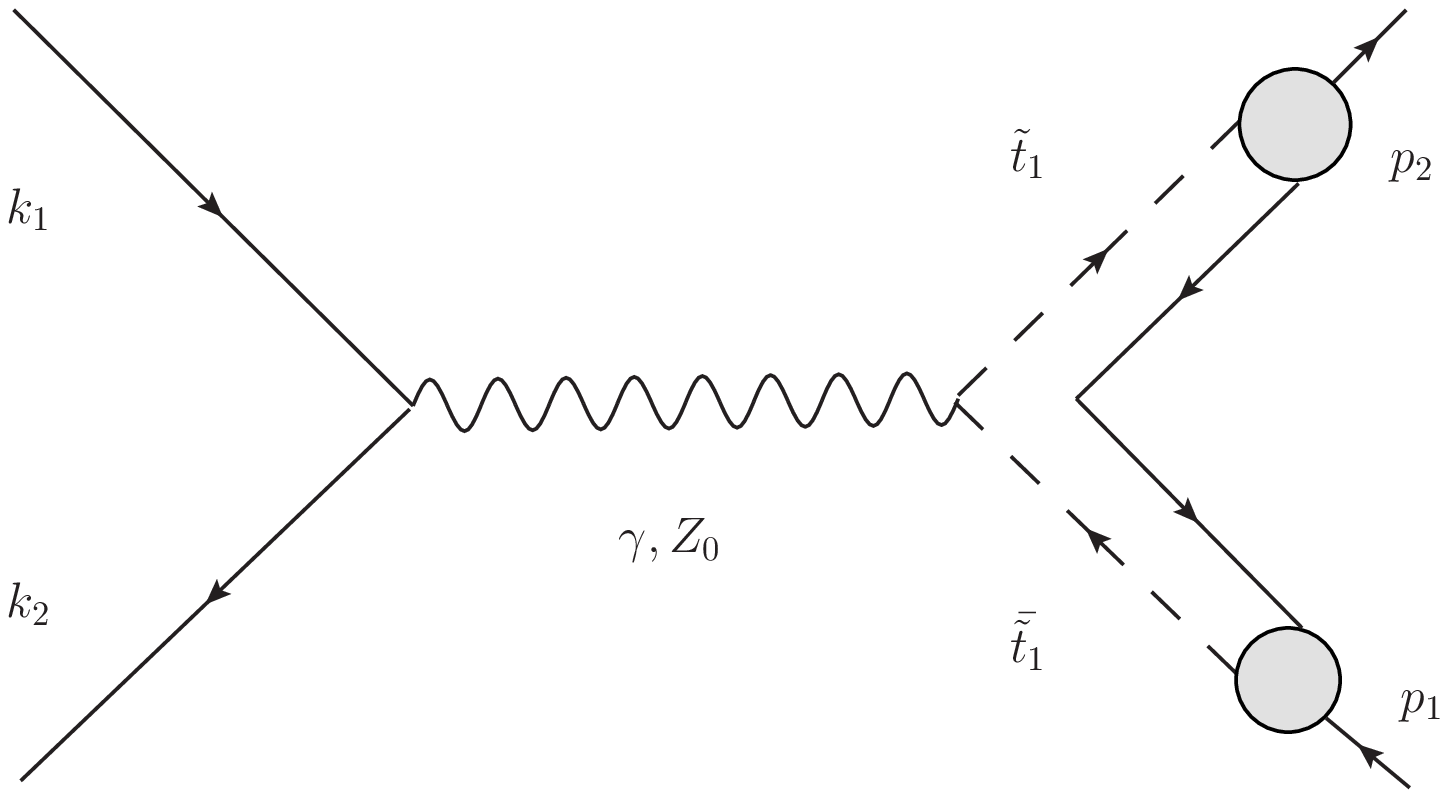}
\caption{The process of $e^+e^-\rightarrow \tilde X\bar{\tilde X}$.}
\label{ststbaryon}
\end{minipage}
\end{figure}

Fig.\ref{bbmeson} and Fig.\ref{ststbaryon} show the leading order
Feynman diagrams for the processes $e^+e^-\rightarrow B\bar B$
and  $e^+e^-\rightarrow \tilde X\bar{\tilde X}$ respectively.
The transition amplitudes, for mesons are
\begin{eqnarray}
i\mathcal M_B=\xi(-\omega)(-i\frac{1}{3}e(p_2-p_1)_\mu)
\frac{-i}{s_1}\bar v(k_2)(-ie\gamma^{\mu})u(k_1)\,,
\end{eqnarray}
and for mesinos are
\begin{eqnarray}\begin{array}{rl}
i\mathcal M_{\tilde X}=&\xi(-\omega)[\bar u(p_2)i\frac{2}{3}e\frac{(p_2-p_1)_\mu}{2 m_{\tilde t_1}}v(p_1)
\frac{-i}{s_2}\bar v(k_2)(-ie\gamma^{\mu})u(k_1)\\
&+\bar u(p_2)g_{tz}\frac{(p_2-p_1)_\mu}{2 m_{\tilde t_1}}v(p_1)
\frac{-i}{s_2-m^2_Z}\bar v(k_2)\gamma_\mu g_{ez}u(k_1)]\,.
\end{array}\end{eqnarray}
Here $\omega=v\cdot v'=\frac{s}{2m^2}-1$, $k_1$ and $k_2$ are the momenta
of the incoming electron and positron, $p_1$ and $p_2$ are the momenta of the
outgoing anti-hadron and hadron. It is noted that the hadronic matrix
elements are determined according to the superflavor symmetry as shown in Eqs.(5) and (6).
Thus we obtain the cross section for pair productions as
\begin{eqnarray}
&\sigma=\frac{1}{2s}\int\frac{d^3p_1}{(2\pi)^3}\frac{1}{2E_1}\frac{d^3p_2}{(2\pi)^3}\frac{1}{2E_2}
(2\pi)^4\delta^4(p_1+p_2-k_1-k_2)\frac{1}{4}\displaystyle {\sum_{\substack{spin}}}|\mathcal M|^2.
\end{eqnarray}
The final expression includes the Isgur-Wise function $|\xi(-\omega)|^2$ which determines the hadronic matrix elements and
manifests the non-perturbative QCD effects in the hadronization. As mentioned above, we cannot use the data
to fix the parameters, so generally we will obtain the values of the Isgur-Wise function for certain $\omega$ by employing some phenomenological
models.

\section{Numerical analysis}

So far, the collider experiments including Tevatron and LHC have not set stringent constraints on $m_{\tilde{t}_1}$ \cite{Kim:2011sv,Belanger:2012mk} yet,
and we would assume $m_{\tilde{t}_1}$
varying from 200 GeV to 500 GeV.

In our numerical calculation, $m_B=5.3$ GeV,  $m_{\tilde{t}_1}=210\sim250$ GeV is taken
for $\sqrt s=500$ GeV and $m_{\tilde{t}_1}=420\sim 500$ GeV for $\sqrt s=1$ TeV respectively, the running Weinberg angle $\sin^2\theta_w$
is taken as $\sin^2\theta_w=0.2398$
for $\sqrt s=500$ GeV and $\sin^2\theta_w=0.2444$ for $\sqrt s=1$ TeV, $\alpha_e$ is approximately equal to $\alpha_e(m_Z)=1/128.78$,
the range of mixing angle $\theta_t$ is uncertain and generally can span in a rather wide range of $0\sim \pi$.
Following Ref.~\cite{Hikasa:1987db}, in our computation we take a few special values of $\cos^2\theta_t$ as 0, $1/2$ and $1$.
Our results obviously depend on the concrete value of $|\xi(-\omega)|^2$. We need to extrapolate $\xi(\omega)$ from a transition region into the
annihilation region as $\omega\to -\omega$, and we can write the Isgur-Wise function as
\begin{equation}
\xi(-\omega)=1-\rho^2(|\omega|-1)+c(|\omega|-1)^2+\cdot\cdot\cdot,
\end{equation}
where the parameters $\rho$ and $c$ are calculated in the lattice QCD \cite{Roy:2012dj}.

Many authors have calculated the numerical value $\xi(\omega)$ in different
ways\cite{Olsson:1995yd,Ahmady:1994ci,Huang:1996hj,Douglas:1999wr,Krutov:2000kt}. In their articles
$\xi(\omega)<1$ when $\omega>1$, and all of their results show that $\xi(1.2)\approx0.8$, $\xi(1.4)\approx0.65$,
$\xi(1.6)\approx0.55$ and $\xi(1.8)\approx0.5$ for the processes $B\rightarrow D$\cite{Olsson:1995yd,Ahmady:1994ci,Huang:1996hj,Douglas:1999wr,Krutov:2000kt}.
A brief discussion about numerical value of the $|\xi(-\omega)|^2$ will be
given in the next section.  In Tab.\ref{stst500} and \ref{stst1t}, we list the production rates of the SUSY mesinos for various $\omega$-values.


\begin{center}
\begin{table}[!htbp]
\begin{tabular}[c]{|l|l|l|l|l|l|}\hline
$\omega$                          &1.00  &1.17   &1.36   &1.58   &1.83\\\hline
$m_{\tilde{t}_1}$(GeV)               &250   &240    &230    &220    &210\\\hline
$\sigma^{expected}(e^+e^-\rightarrow \tilde X\bar{\tilde X})$(fb)($\cos^2\theta_t=0$)
                                  & 0     & 0.34 & 1.33 & 2.73 & 4.77\\\hline
$\sigma^{expected}(e^+e^-\rightarrow \tilde X\bar{\tilde X})$(fb)($\cos^2\theta_t=1/2$)
                                  & 0     & 0.33 & 1.28 & 2.64 & 4.61 \\\hline
$\sigma^{expected}(e^+e^-\rightarrow \tilde X\bar{\tilde X})$(fb)($\cos^2\theta_t=1$)
                                  & 0     & 0.50 & 1.93 & 3.97 & 6.94\\\hline
\end{tabular}
\caption{The cross sections of $\sigma(e^+e^-\rightarrow \tilde X\bar{\tilde X})$
with the center of mass energy $\sqrt{s}$=500 GeV.
\label{stst500}}
\end{table}
\end{center}
\begin{center}
\begin{table}[!htbp]
\newcommand{\tabincell}[2]{\begin{tabular}{@{}#1@{}}#2\end{tabular}}
\begin{tabular}[c]{|l|l|l|l|l|l|}\hline
$\omega$                          &1.00  &1.17   &1.36   &1.58   &1.83   \\\hline
$m_{\tilde{t}_1}$(GeV)               &500   &480    &460    &440    &420    \\\hline
$\sigma^{expected}(e^+e^-\rightarrow \tilde X\bar{\tilde X})$(fb)($\cos^2\theta_t=0$)
                                  & 0     & 0.08 & 0.34 & 0.69 & 1.21\\\hline
$\sigma^{expected}(e^+e^-\rightarrow \tilde X\bar{\tilde X})$(fb)($\cos^2\theta_t=1/2$)
                                  & 0     & 0.08 & 0.32 & 0.65 & 1.14 \\\hline
$\sigma^{expected}(e^+e^-\rightarrow \tilde X\bar{\tilde X})$(fb)($\cos^2\theta_t=1$)
                                  & 0     & 0.12 & 0.46 & 0.95 & 1.66\\\hline
\end{tabular}
\caption{The cross section of $\sigma(e^+e^-\rightarrow \tilde X\bar{\tilde X})$
with the center of mass energy $\sqrt{s}$=1 TeV.
\label{stst1t}}
\end{table}
\end{center}

\begin{center}
\begin{table}[!htbp]
\begin{tabular}[c]{|l|l|l|l|l|l|}\hline
$\omega$                           &1.00  &1.17  &1.36  &1.58  &1.83  \\\hline
$\sqrt{s}$(GeV)                    &10.60 &11.04 &11.52 &12.04 &12.62 \\\hline
$\sigma(e^+e^-\rightarrow B\bar B)$(pb)
                                   & 0     & 0.94 & 1.58 & 1.84 & 2.06 \\\hline
\end{tabular}
\caption{The cross section of $\sigma(e^+e^-\rightarrow B\bar B)$ for the CM energy $\sqrt s$ of the B-factories.
\label{bb500}}
\end{table}
\end{center}

In Tab.\ref{stst500} and \ref{stst1t} we show the numerical values
of the cross sections in the range of $m_{\tilde{t}_1}=250\sim210$ GeV and $m_{\tilde{t}_1}=500\sim420$ GeV  corresponding
to $\omega$ varying from $1$ to $1.83$ at the center of mass energy $\sqrt s=500$ GeV and $\sqrt s=1$ TeV respectively.
Tab.\ref{bb500} gives the results of $\sigma(e^+e^-\rightarrow B\bar B)$ with the same
$\omega$ values as that in Tabs. \ref{stst500}, \ref{stst1t}.

\begin{center}
\begin{table}[!htbp]
\begin{tabular}[c]{|l|l|l|l|l|l|}\hline
$\omega$                          &1.00  &1.17   &1.36   &1.58   &1.83\\\hline
$m_{\tilde{t}_1}$(GeV)               &250   &240    &230    &220    &210\\\hline
$\sigma^{theor}(e^+e^-\rightarrow \tilde{t}_1\bar{\tilde{t}}_1)$(fb)($\cos^2\theta_t=0$)
                                  & 0     & 3.14 & 8.62 &15.34 &22.86\\\hline
$\sigma^{theor}(e^+e^-\rightarrow \tilde{t}_1\bar{\tilde{t}}_1)$(fb)($\cos^2\theta_t=1/2$)
                                  & 0     & 3.04 & 8.33 &14.84 &22.12 \\\hline
$\sigma^{theor}(e^+e^-\rightarrow \tilde{t}_1\bar{\tilde{t}}_1)$(fb)($\cos^2\theta_t=1$)
                                  & 0     & 4.57 &12.54 &22.32 &33.28\\\hline
\end{tabular}
\caption{The cross section of $\sigma(e^+e^-\rightarrow \tilde{t}_1\bar{\tilde{t}}_1)$
with the center of mass energy $\sqrt{s}$=500 GeV.
\label{ststq500}}
\end{table}
\end{center}
\begin{center}
\begin{table}[!htbp]
\newcommand{\tabincell}[2]{\begin{tabular}{@{}#1@{}}#2\end{tabular}}
\begin{tabular}[c]{|l|l|l|l|l|l|}\hline
$\omega$                          &1.00  &1.17   &1.36   &1.58   &1.83   \\\hline
$m_{\tilde{t}_1}$(GeV)               &500   &480    &460    &440    &420    \\\hline
$\sigma^{theor}(e^+e^-\rightarrow \tilde{t}_1\bar{\tilde{t}}_1)$(fb)($\cos^2\theta_t=0$)
                                  & 0     & 0.79 & 2.18 & 3.87 & 5.77\\\hline
$\sigma^{theor}(e^+e^-\rightarrow \tilde{t}_1\bar{\tilde{t}}_1)$(fb)($\cos^2\theta_t=1/2$)
                                  & 0     & 0.75 & 2.06 & 3.67 & 5.47 \\\hline
$\sigma^{theor}(e^+e^-\rightarrow \tilde{t}_1\bar{\tilde{t}}_1)$(fb)($\cos^2\theta_t=1$)
                                  & 0     & 1.09 & 2.99 & 5.32 & 7.94\\\hline
\end{tabular}
\caption{The cross section of $\sigma(e^+e^-\rightarrow \tilde{t}_1\bar{\tilde{t}}_1)$
with the center of mass energy $\sqrt{s}$=1 TeV.
\label{ststq1t}}
\end{table}
\end{center}
In Tab.\ref{ststq500} and \ref{ststq1t} we also list the cross sections of the
process $e^+e^-\rightarrow \tilde{t}_1\bar{\tilde{t}}_1$ with the $m_{\tilde{t}_1}$
varies in the range of $250\sim210$ GeV and $500\sim420$ GeV.
The authors of Ref.\cite{Bartl:2000kw} calculated the cross section and gave its dependence on the CM energy of ILC, while assuming 
$m_{\tilde{t}_1}$ to be 200 GeV and 420 GeV respectively. Our results are generally consistent 
with theirs. From the data above we
can find that the ratio of a scalar top quark pair transiting into a SUSY mesino pair
is about $10\%\sim20\%$.

\section{Conclusion and Discussion}

With the help of superflavor symmetry, we associate the production of
the stop-mesino pairs with $B\bar B$ neat their thresholds. Thus the
production rate of the SUSY mesino pair near its production threshold at the future ILC
can be compared with the B-meson pair production rate at the B-factories.
However, the experimental measurement on the continuum contribution to $B\bar B$ at the B-factory is not available
because it is buried in large background corresponding to various resonances
which make extraction of the continuum contribution not reliable.

So we use the superflavor symmetry where the non-perturbative QCD effects are included in a unique Isgur-Wise function
$\xi(|\omega|)$ to analyze  the mesino production directly. Meanwhile in the same scheme, we also calculate
the production rate of $B\bar B$ neat its production threshold. The obtained rate is nothing but the continuum
contribution to the process $e^+e^-\to B\bar B$, and it is a by-product of this research.

From Ref.\cite{Olsson:1995yd,Ahmady:1994ci,Huang:1996hj,Douglas:1999wr,Krutov:2000kt}
we can find $\xi(\omega)$ decrease with $\omega$, so when $\omega$ increases,
the value of $|\xi(\omega)|^2$ is less than $1$. Therefore, the real production
rate of the mesino pair is slightly less than the value we list in Tab.\ref{stst500} and \ref{stst1t}.
On the other hand, the heavy quark/squark pair captures a light quark pair from vacuum to
form a meson/mesino pair. It means that as the velocity of the heavy quark/squark pair increases,
the probability of capturing a light quark pair from vacuum decreases,
thus when $\omega$ increases, $|\xi(\omega)|^2$ decreases from $1$.



The ILC is proposed to begin running in 10 years.
Its early stage is designed to be running at the center of mass energy  of $\sqrt s=500$ GeV
with yearly integrated luminosity $500$\ fb$^{-1}$, then the energy will be updated to $1$ TeV with the integrated luminosity
$1000$\ fb$^{-1}$ \cite{Baer:2013cma}. In Tab.\ref{stste500} and Tab.\ref{stste1t} we list the numbers of the SUSY
stop mesino pairs generated per year at ILC for $\sqrt s=500$ GeV
and $\sqrt s=1$ TeV respectively.

\begin{center}
\begin{table}[!htbp]
\begin{tabular}[c]{|l|l|l|l|l|l|}\hline
$\omega$                          &1.00  &1.17   &1.36   &1.58   &1.83\\\hline
$m_{\tilde{t}_1}$(GeV)               &250   &240    &230    &220    &210\\\hline
events                           & 0     & 170 & 665 &1365 &2385\\\hline
events
                                  & 0     & 165 & 640 &1320 &2305 \\\hline
events
                                  & 0     & 250 & 965 &1985 &3470\\\hline
\end{tabular}
\caption{The number of the event predicted in ILC with the center of mass energy $\sqrt{s}$=500 GeV
and luminosity $500$\ fb$^{-1}$.
\label{stste500}}
\end{table}
\end{center}
\begin{center}
\begin{table}[!htbp]
\newcommand{\tabincell}[2]{\begin{tabular}{@{}#1@{}}#2\end{tabular}}
\begin{tabular}[c]{|l|l|l|l|l|l|}\hline
$\omega$                          &1.00  &1.17   &1.36   &1.58   &1.83   \\\hline
$m_{\tilde{t}_1}$(GeV)               &500   &480    &460    &440    &420    \\\hline
events
                                  & 0     &  80 & 340 & 690 & 1210\\\hline
events
                                  & 0     &  80 & 320 & 650 & 1140 \\\hline
events
                                  & 0     & 120 & 460 & 950 & 1660\\\hline
\end{tabular}
\caption{The number of the event predicted in ILC with the center of mass energy $\sqrt{s}$=1 TeV
and luminosity $1000$\ fb$^{-1}$.
\label{stste1t}}
\end{table}
\end{center}
Taking into account of detection efficiency there would be
a sufficiently large amount of events to be observed.

Following suggestions given in literature, we consider the scalar top quark $\tilde{t}_1$ as the NLSP, thus
the mesino which consists of $\tilde{t}_1$ and a SM anti-quark has very distinctive characters. It is a
fermion of baryon number being zero, so it is completely different from the SM baryons. Moreover, as  R-parity is conserved,
the main decay mode of stop is $\tilde{t}_1\to \tilde{\chi}_1^0$+ SM quark(+others) where $\tilde{\chi}_1^0$ is
the lightest SUSY particle (LSP): the neutralino.
If the mass splitting between stop and neutralino is sufficiently small, the decay channel $\tilde{t}_1\to \tilde{\chi}_1^0+b+W^{(*)}$ is restricted by the final state phase space,
Another probable channel would be $\tilde{t}_1\to \tilde{\chi}_1^0+c (u)$ which occur via loops.
so that it is suppressed. The main decay mode of the mesino is via the process where $\tilde{t}_1$ transits to
$\tilde{\chi}_1^0$ by radiating a SM quark which later
combines with the constituent anti-quark (as a spectator) in the mesino to constitute  a SM meson (either pseudoscalar or vector).
Thus observable process is that a
fermion of $B=0$ transiting to a SM meson plus missing energy. This signal is very clean and unique,
so that from such signal, one can immediately identify
the SUSY mesino. Since the stop mesino can be charged ($\tilde{t}_1$+$\bar d (or\; \bar s)$),
one cannot miss its trajectory.

Therefore we expect a stop mesino with a relative long
lifetime to be detected at the facilities which will be available in the not-far future.
The authors of Ref. \cite{Kim:2014yaa} also suggested that the stoponium may be observed via
its decay products $\gamma\gamma$ and $ZZ$ at LHC in the following 14 TeV running.
Definitely, they are more easily to be observed at ILC due to its clean background.

Our numerical computations depend on the the Isgur-Wise function which manifests the non-perturbative QCD effects. Since
the function is phenomenologically introduced it brings up uncertainties to our numerical results. As we expected, if the continuum
contribution to $e^+e^-\to B\bar B$ could be extracted from the data or directly experimentally measured, we would be able
to greatly reduce the theoretical uncertainties and help to draw definite conclusion.

It is also noted that the updated SUSY hadron search results given by CMS \cite{Chatrchyan:2012dxa}
and ATLAS \cite{Aad:2013gva} Collaborations indicate that
SUSY hadrons' lifetimes should be shorter than $\mu$'s if they exist with  sub-TeV masses. Indeed, if their lifetimes are too short,
it is disadvantageous for their detection, but there still is possibility for direct detection of stop mesinos.
We lay hope on the next run of LHC, which may provide information about the SUSY particles,
and look forward to the future ILC, where the SUSY particles can be better identified. Moreover, the proposed CEPC (Circular electron-proton
collider) and the tera Z-factory in China might also join the project for searching mesinos.

\section*{Acknowledgments}

This  work is supported by National Natural Science Foundation of China under the
Grant Number 11075079, 11135009 and 11005061.

\end{document}